\documentclass[emulateapj]{aastex631}

\usepackage{amssymb}
\usepackage{amsmath}
\usepackage{gensymb}
\usepackage{graphicx}
\usepackage{xcolor}
\usepackage{natbib}

\newcommand{\hi}{H{\sc i}}
\newcommand{\himf}{H{\sc i}MF}
\newcommand{\mh}{\rm M_{H{\textsc i}}}
\newcommand{\MHI}{\rm M_{H{\textsc i}}}
\newcommand{\mb}{\rm M_B}
\newcommand{\hii}{H{\sc i} 21\,cm}
\newcommand{\msun}{\rm M_{\odot}}

\newcommand{\kms}{km~s$^{-1}$}

\shorttitle{The H{\sc i} mass function at $z \sim 0.35$}
\shortauthors{Bera et al.}

\begin{document}

\title{{\Large The H{\sc i} mass function of star-forming galaxies at $\mathbf{z \sim 0.35}$}}

\correspondingauthor{Nissim Kanekar}
\email{nkanekar@ncra.tifr.res.in}

\author{Apurba Bera}
\affil{Inter-University Centre for Astronomy and Astrophysics, Pune 411007, India}
\affil{National Centre for Radio Astrophysics, Tata Institute of Fundamental Research, Pune 411007, India}

\author{Nissim Kanekar}
\affil{National Centre for Radio Astrophysics, Tata Institute of Fundamental Research, Pune 411007, India}

\author{Jayaram N. Chengalur}
\affil{National Centre for Radio Astrophysics, Tata Institute of Fundamental Research, Pune 411007, India}

\author{Jasjeet S. Bagla}
\affil{Indian Institute of Science Education and Research Mohali, Knowledge City, Sector 81, Sahibzada Ajit Singh Nagar, Punjab 140306, India}


\begin{abstract}

The neutral atomic hydrogen (H{\sc i}) mass function (H{\sc i}MF) describes the distribution of the H{\sc i} content of galaxies at any epoch; its evolution provides an important probe of models of galaxy formation and evolution. Here, we report Giant Metrewave Radio Telescope H{\sc i} 21cm spectroscopy of blue star-forming galaxies at $z\approx0.20-0.42$ in the Extended Groth Strip, which has allowed us to determine the scaling relation between the average H{\sc i} mass ($\rm{M_{H{\textsc i}}}$) and the absolute B-band magnitude ($\rm{M_B}$) of such galaxies at $z \approx 0.35$, by stacking the H{\sc i} 21cm emission signals of galaxy subsamples in different $\rm{M_B}$ ranges.  We combine this $\rm{M_{H{\textsc i}}-M_B}$ scaling relation (with a scatter assumed to be equal to that in the local Universe) with the known B-band luminosity function of star-forming galaxies at these redshifts to determine the H{\sc i}MF at $z\approx0.35$. We show that the use of the correct scatter in the $\rm{M_{H{\textsc i}}-M_B}$ scaling relation is critical for an accurate estimate of the H{\sc i}MF. We find that the H{\sc i}MF has evolved significantly from $z\approx0.35$ to $z\approx0$, i.e. over the last four Gyr, especially at the high-mass end. High-mass galaxies, with $\rm{M_{H{\textsc i}}\gtrsim10^{10}\  M_\odot}$, are a factor of $\approx3.4$ less prevalent at $z\approx0.35$ than at $z \approx 0$. Conversely, there are more low-mass galaxies, with  $\rm{M_{H{\textsc i}} \approx10^9\ {M}_\odot}$, at $z\approx0.35$ than in the local Universe. While our results may be affected by cosmic variance, we find that massive star-forming galaxies have acquired a significant amount of H{\sc i} through merger events or accretion from the circumgalactic medium over the past four Gyr. 

\end{abstract}


\section{Introduction} \label{sec:intro}

The neutral atomic hydrogen (\hi) mass function (\himf) of galaxies, the number of galaxies with a given \hi\ mass per unit cosmic volume, gives the distribution of \hi\ across galaxies in the Universe and is a fundamental descriptor of galaxy populations. Observational constraints on the redshift evolution of the \himf\ are essential to test the predictions of different galaxy evolution models \citep[e.g.][]{ somerville99mnras, cole00mnras, bahe16, dave17, dave2020}. This is especially the case in the modern view of galaxies in which the baryon cycle, the exchange of gas between a galaxy and its circumgalactic medium (CGM), plays a critical role in galaxy evolution \citep[e.g.][]{peroux2020}. However, the \himf\ has so far only been determined at the present epoch, $z \approx 0$ \citep[e.g.][]{zwaan05mnras,jones18mnras}; the lack of observational constraints beyond the local Universe has meant that its redshift evolution is entirely unknown. A direct measurement of the \himf\ at cosmological distances requires the detection of \hii\ emission from a large number of individual galaxies at the redshift of interest. Unfortunately, this is very difficult to accomplish with present-day radio telescopes, due to the low Einstein-A coefficient of the hyperfine \hii\ transition. Indeed, only a single galaxy has so far been detected in \hii\ emission at $z \gtrsim 0.3$ \citep[e.g.][]{fernandez16apj}. 

Over the last few decades, several studies have found that the \hi\ mass ($\MHI$) of galaxies in the local Universe correlates with their optical luminosity \citep[e.g.][]{haynes84aj, toribio11apj2}; the tightest correlation is between $\MHI$ and the absolute B-band magnitude, $\mb$ \citep{denes14mnras}. \citet{briggs90aj} pointed out that such a scaling relation between \hi\ mass and $\mb$ could be combined with the B-band luminosity function of galaxies to determine the \himf. Indeed, this was the approach originally used to determine the \himf\ at $z \approx 0$, before the era of unbiased wide-field \hii\ emission surveys \citep{rao93apj,zwaan01mnras}. Deep optical photometry has yielded accurate B-band luminosity functions for large samples of galaxies out to high redshifts, $z \gtrsim 1$ \citep[e.g.][]{willmer06apj,lopezsanjuan17aa}. A measurement of the $\MHI - \mb$ relation in galaxies at cosmological distances might then be combined with the B-band luminosity function to measure the \himf\ at these epochs.

Although it is difficult to detect \hii\ emission from individual galaxies at $z \gtrsim 0.2$ with current radio telescopes, the method of \hii\ emission ``stacking'' allows us to measure the average \hi\ mass and average \hi\ properties of galaxy samples at cosmological distances \citep[e.g.][]{zwaan00thesis, chengalur01aa, lah07mnras, delhaize13mnras, rhee16mnras, rhee18mnras, kanekar16apjl, bera19apjl, chowdhury20nature, chowdhury21apjl, chowdhury22apjl, chowdhury22survey, chowdhury22baryon, sinig22random}. Such \hii\ stacking experiments also allow one to determine galaxy scaling relations, the average Tully-Fisher relation, etc \citep[e.g.][]{fabello11,fabello12,meyer16}.

We have used the Giant Metrewave Radio Telescope to carry out a deep, $\approx 350$-hr, \hii\ emission survey of the Extended Groth Strip \citep[EGS;][Bera et al., in prep.]{bera19apjl}. In this {\it Letter}, we present the first measurement of the $\MHI - \mb$ relation in star-forming galaxies at cosmological distances, $z \approx 0.35$. We further use this relation along with the B-band luminosity function of galaxies at $z \approx 0.2-0.4$ to obtain the first determination of the \himf\ at cosmological distances. Throughout this work, we use a flat $\Lambda$-cold dark matter ($\Lambda$CDM) cosmology, with ($\rm H_0$, $\rm \Omega_{m}$, $\rm \Omega_{\Lambda})=(70$~km~s$^{-1}$~Mpc$^{-1}$, $0.3, 0.7)$. All magnitudes are in the AB system \citep{oke74apjs}.

\section{Observations and data analysis} \label{sec:data}

We used the Band-5 receivers of the upgraded GMRT \citep{gupta17ugmrt} to observe the EGS over four observing cycles between March 2017 and June 2019, in proposals 31\_038 (P.I. J.~S.~Bagla), 34\_083 (P.I. N.~Kanekar), 35\_085 (P.I. A.~Bera), and 36\_064 (P.I. A.~Bera). The total observing time was $\approx 347$~hours, with a total on-source time of $\approx 250$~hours. The observations and data analysis are described in detail by \citet{berainprep}; a brief summary is provided below. 

Our GMRT observations of the EGS covered the frequency range $970 - 1370$~MHz, using the GMRT Wideband Backend (GWB) as the correlator, with a bandwidth of 400~MHz sub-divided into 8,192 spectral channels. The basic data editing, to remove non-working antennas and data affected by systematic effects such as radio frequency interference (RFI), and calibration to determine the antenna-based gains and bandpasses were carried out in the classic {\sc aips} package \citep{greisen03book}. For each observing cycle, the calibrated EGS visibilities were then combined into a single data file. A self-calibration and data editing procedure was run on each data set, involving a few rounds of imaging and phase self-calibration, followed by a couple of rounds of imaging and amplitude-and-phase self-calibration, and data editing. This process was continued until both the antenna-based gains and the residual visibilities did not improve on further self-calibration and data editing. Following this, all detected continuum emission was subtracted from the calibrated spectral-line visibilities, using the {\sc aips} task {\sc uvsub}. The residual visibilities were then imaged using the task {\sc tclean} in the {\sc casa} package \citep[version 5.6;][]{mcmullin07book} to produce spectral-line data cubes in the barycentric frame; this imaging was carried out using w-projection \citep{cornwell08} and Briggs weighting, with {\sc robust}=0.5 \citep{briggs95}. Each observing cycle yielded one independent spectral-line data cube. A frequency-dependent correction for the shape of the antenna primary beam was then applied to each cube. The final spectral cubes have a frequency resolution of 97.7~kHz, corresponding to a velocity resolution of $\approx 21-30$~\kms\ across the observing band. The synthesized beams of the cubes have FWHMs of $\approx 2\farcs6 - 3\farcs3$, corresponding to spatial resolutions of $\approx 9 - 18$~kpc for the redshift range $z = 0.20-0.42$. 

\section{\hii~spectral line stacking} \label{sec:stacking}

\subsection{Sample selection and the \hii\ subcubes} \label{subsec:sample}

The EGS was chosen as the target of our \hii\ emission survey due to the availability of accurate spectroscopic redshifts for a large number of galaxies with ${\rm R_{AB}} \leq 24.1$ at $z \approx 0.20-0.42$, from the DEEP2 and DEEP3 Galaxy Surveys \citep{cooper12mnras, newman13apjs}. The redshift accuracy is sufficient \citep[velocity errors~$\lesssim 62$~\kms;][]{newman13apjs} to allow stacking of the \hii\ emission signals of the individual galaxies, without dilution of the stacked \hii\ signal \citep[e.g.][]{elson2019}. The DEEP2 and DEEP3 redshift coverage implies the limit $z \geq 0.2$ on our sample, while the frequency coverage of the GMRT Band-5 receivers gives the limit $z \lesssim 0.42$; our stacking analysis was hence restricted to galaxies with $0.20 \leq z \leq 0.42$. We further restricted our sample to galaxies (1)~lying within the half-power point of the GMRT primary beam at each galaxy's redshifted \hii\ line frequency, (2)~with reliable redshifts, with redshift quality code, Q~$\geq 3$ \citep{newman13apjs}, and (3)~with absolute blue magnitude $\mb \leq -16$. This yielded a sample of 808 galaxies. The presence of an active galactic nucleus (AGN) in a galaxy could affect its average \hi\ properties; we hence aimed to exclude galaxies containing AGNs from the sample. AGNs have been shown to dominate the population of radio sources with rest-frame 1.4~GHz radio luminosity $L_{\rm 1.4 \; GHz} \gtrsim 2 \times 10^{23}$~W~Hz$^{-1}$ \citep{condon02aj,smolcic08apj}. We applied this radio luminosity threshold to our continuum image of the EGS to identify, and then exclude, possible AGNs from our sample. In addition, galaxies identified as AGN hosts in the DEEP2 or DEEP3 catalogues, based on their optical spectra, were also excluded from the sample. A total of 84 AGNs were identified and excluded by these criteria. Next, using the DEEP2 colour criterion ${\rm U-B}+0.032 \times ({\rm M_B}+21.62)-1.035<0$ \citep[where U and B are the absolute magnitudes in the rest-frame U and B bands; ][]{willmer06apj, coil08apj}, we excluded 101 red galaxies, restricting our sample to blue star-forming galaxies at $0.20 \leq z \leq 0.42$. 

For each target galaxy in our sample, we extracted a subcube 
centred at the position and redshifted \hii\ line frequency of the galaxy from each of the final four spectral cubes of the four observing cycles. The four subcubes of each target galaxy were treated independently in the analysis, so that any systematic errors in one of the observing cycles would not cause all data on the galaxy to be dropped from our sample \citep[e.g.][]{chowdhury22survey,berainprep}. The subcubes were converted from flux density (in Jy) to luminosity density (in Jy~Mpc$^2$), using the relation $L_{{\rm H}{\textsc i}} = {4 \pi . F_{\nu}. d_L^2(z)}/(1+z)$, where $d_L(z)$ is the luminosity distance and $F_{\nu}$ is the flux density.

Each subcube was convolved with a Gaussian beam to a spatial resolution of 30~kpc at the galaxy redshift, and then appropriately normalized to ensure that the peak of the convolved (with the same kernel) point spread function is unity. The spatial resolution of 30~kpc was chosen to ensure that the stacked \hii\ emission signal is not resolved: smoothing the subcubes to coarser resolutions yielded a total \hii\ emission signal consistent with that obtained at a resolution of 30~kpc \citep{berainprep}. Each subcube was also interpolated to a velocity resolution of 30~\kms, and regridded to a spatial pixel size of 5~kpc at the galaxy redshift. A second-order polynomial was then fitted to the spectrum at each spatial pixel of each subcube, excluding the central $\pm 200$~\kms, and subtracted out. The residual subcubes have spatial extents of $500 \ {\rm kpc} \times 500 \ {\rm kpc}$, and cover the velocity range $\pm 1000$~\kms, at the galaxy redshift. 

Finally, a set of statistical tests were used to test the subcubes for the presence of systematic effects arising due to, e.g., RFI, deconvolution errors, etc \citep[see][for details]{berainprep}. Subcubes that were identified as being affected by such systematics were excluded from the sample. Additionally, to mitigate the possible effect of source confusion, 14 galaxies with `neighbours' were excluded from the sample. A `neighbour' is defined as any galaxy with $\mb \leq -16$ within a spatial resolution element ($30$~kpc) of the target galaxy  and within $\pm$ 300 \kms~in redshift. This yielded a final sample of 464 unique galaxies, with 1665 subcubes, for our \hii\ stacking experiment. The average redshift of the sample is $\langle z \rangle = 0.35$.

\subsection{Stacking the \hii\ emission} \label{subsec:method}

To determine the $\MHI - M_B$ relation at $z \sim 0.35$, we first divided the galaxy sample into three independent $\mb$ bins. The number of $\mb$ bins and the bin widths were selected to ensure that the average \hii\ signal is detected at $> 4\sigma$ significance in each bin\footnote{We note that slight changes in the ranges of the three $\mb$ bins were found to have no significant effect on the inferred $\MHI - \mb$ scaling relation.}. The $\mb$ ranges of the three bins are listed in the first column of Table~\ref{table:mhi-mb}.

The stacking of the \hii\ emission signals was carried out  independently for each $\mb$ bin. For each bin, the subcubes of all galaxies in the bin were stacked, plane-by-plane, to produce a stacked spectral cube. 
The stacking was done without assigning any weights to the different subcubes, i.e. all of them were given equal weights. This was done because weighting based on the RMS noise in flux density units would give high weights to the relatively few galaxies in the central regions of the field. Conversely, weighting based on the RMS noise in luminosity density units would give high weights to galaxies at the lowest redshifts. Further, the errors on the final HI masses are dominated by the jackknife errors (see Table~\ref{table:mhi-mb}), rather than by the errors on the stacked \hii\ emission signals. Given that the lack of weights simplifies the interpretation of the stacked emission signal, we chose to not use weights in the stacking process. 
A second-order polynomial was then fitted to the spectrum at each spatial pixel of the cubes, excluding the central $\pm 200$~\kms, and subtracted out. Finally, each residual cube was Hanning-smoothed to, and resampled at, a velocity resolution of 60~\kms, to obtain the final stacked \hii\ spectral cube for each bin. 

\begin{figure*}[t!]
\begin{center}
\includegraphics[scale=0.58, trim={0.4cm 0cm 0.3cm 0.2cm}, clip]{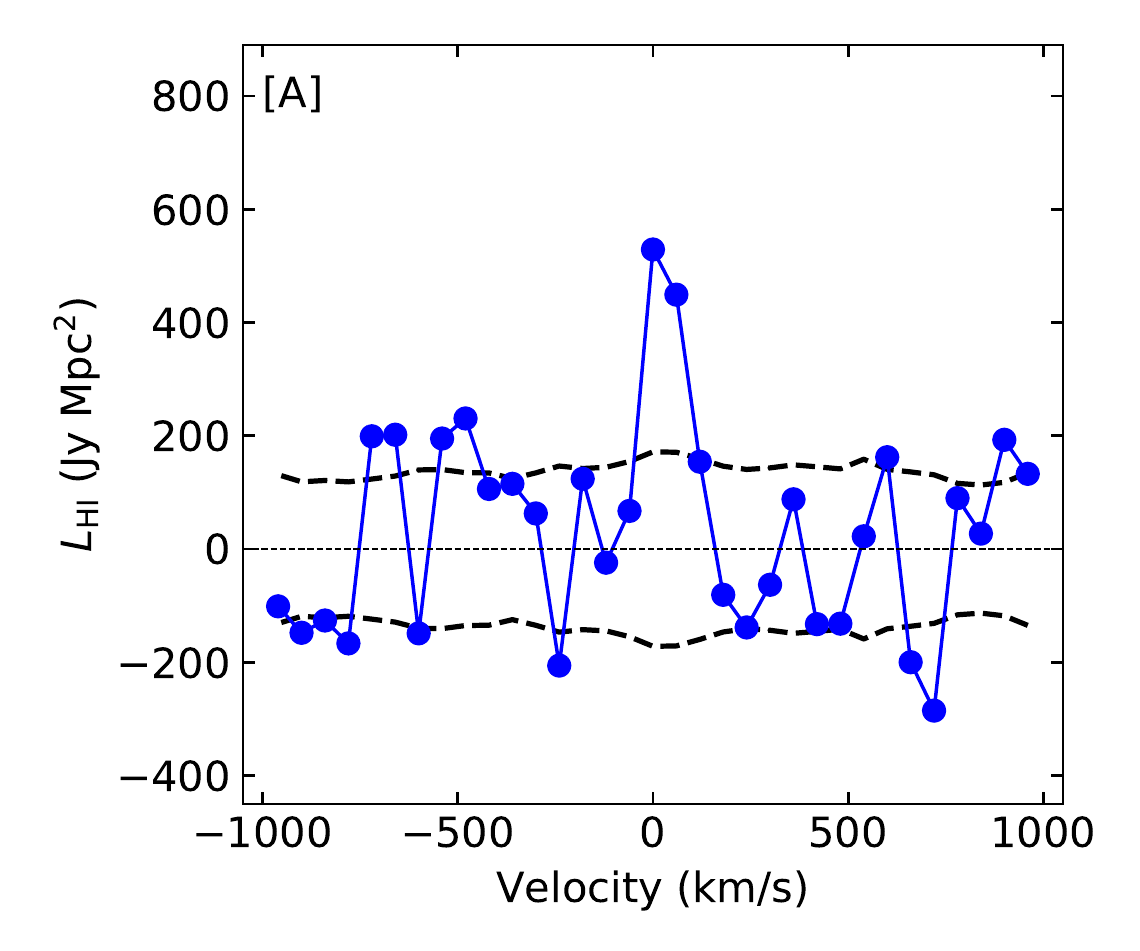}
\includegraphics[scale=0.58, trim={1.4cm 0cm 0.3cm 0.2cm}, clip]{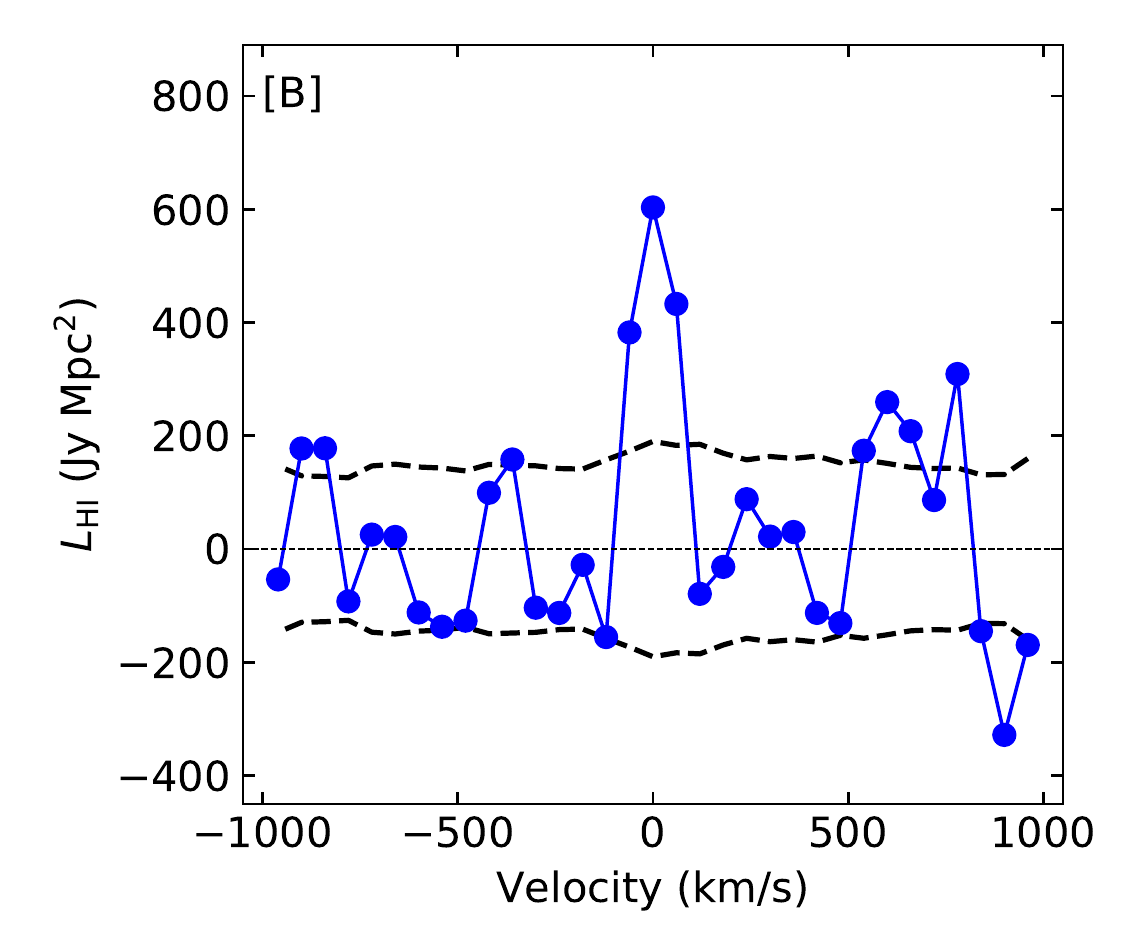}
\includegraphics[scale=0.58, trim={1.4cm 0cm 0.3cm 0.2cm}, clip]{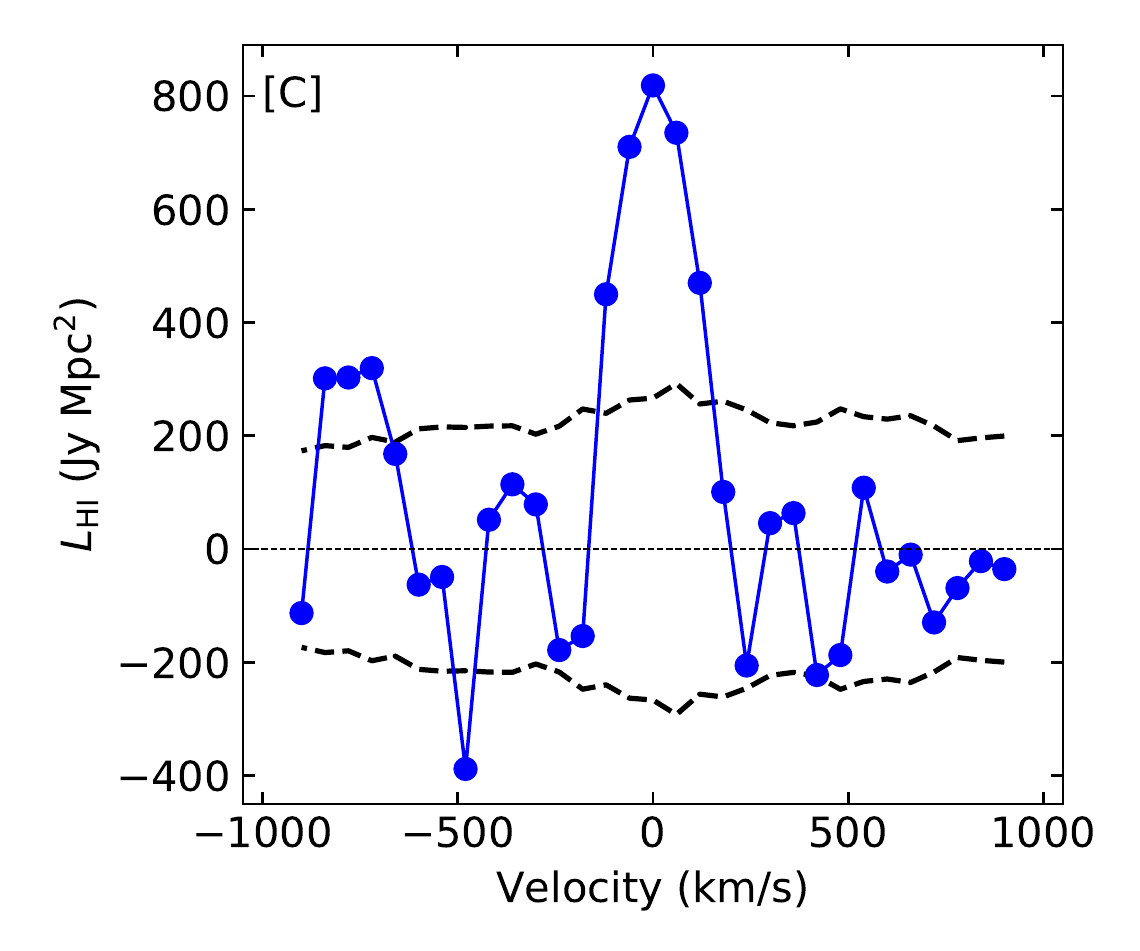}
\caption{The stacked \hii\ emission spectra (in luminosity density) for blue galaxies in the three $\mb $ bins, [A]~$-16.0 \geq \mb > -17.6$, [B]~$-17.6 \geq \mb > -19.2$, and [C]~$-19.2 \geq \mb \geq -22.0$. The dashed black curves in each panel show the RMS noise in the corresponding velocity planes of the stacked spectral cubes. See text for details.}
\label{fig:spectra}
\end{center}
\end{figure*}

The stacked \hii\ spectrum for each $\mb$ bin was obtained by taking cuts through the central pixel of the three final stacked cubes. The final stacked spectra for the three $\mb$ bins are shown in Figure~\ref{fig:spectra}. In all cases, the stacked \hii\ emission signal is detected at $> 4\sigma$ significance. For each bin, the average \hii\ line luminosity was determined by integrating the stacked \hii\ spectrum over all contiguous central spectral channels with $\geq 1.5\sigma$ significance. The RMS noise in the planes of the corresponding stacked cube were used as the measurement errors to estimate the detection significance of the average \hii\ signal for each bin. The velocity-integrated stacked \hii\ line luminosities are listed in the penultimate column of Table~\ref{table:mhi-mb}.

Finally, the stacked \hii\ line luminosities were converted to the average \hi\ mass of the galaxies in each $\mb$ bin via the relation
$\MHI  = 1.86 \times 10^{4} \; \int L_{{\rm H}{\textsc i}}\: dv$, where $\MHI$ is in $\rm M_\odot$ and $\rm \int L_{{\rm H}{\textsc i}}\: dv$ is in Jy~Mpc$^2$~km s$^{-1}$. Jackknife re-sampling was used to estimate the uncertainties on the average \hi\ masses; these are larger than the measurement errors, and include contributions from both sample variance and any underlying systematic effects. The final average \hi\ masses and errors are listed in the last column of Table~\ref{table:mhi-mb}.

We also stacked the \hii\ emission from all 464 galaxies of our sample \citep[][]{berainprep}. The mean \hii\ line luminosity of the full sample is $(8.4 \pm 1.2) \times 10^4 $~Jy~Mpc$^2$~km s$^{-1}$ (measurement errors), while the corresponding mean \hi\ mass is $(1.57 \pm 0.26) \times 10^9 \; \msun $ (jackknife errors). We also measured the median \hi\ mass of the full sample using a median-stacking approach\footnote{Median-stacking refers to taking the median (instead of the mean) of the corresponding pixels, plane-by-plane, of the individual subcubes.}. This yielded a median \hii\ line luminosity of $(6.3 \pm 1.0) \times 10^4 $~Jy~Mpc$^2$~km s$^{-1}$ (measurement errors); the corresponding median \hi\ mass is $(1.17 \pm 0.22) \times 10^9\: \msun $ (jackknife errors). 

\begin{table*}
\begin{center}
\caption{\textbf{The average \hi~mass of blue galaxies in different $\mb $ bins.} }
\begin{tabular}{|c|c|c|c|c|}
\hline
$\mb $ range & Number of galaxies & Median $\mb $ & $ \int L_{{\rm H}{\textsc i}}\: dv$ & $\langle \MHI \rangle$ \\
                &     &  & $10^5\: {\rm Jy\ Mpc^2\ km\ s^{-1}}$ & $10^9\:{\rm M}_\odot$ \\
\hline
$(-17.6,-16.0]$ & 190 & $-17.03$ & $0.59 \pm 0.14$ & $1.09 \pm 0.29$ \\
$(-19.2,-17.6]$ & 195 & $-18.24$ & $0.85 \pm 0.19$ & $1.59 \pm 0.43$ \\
$[-22.0,-19.2]$ & 79  & $-19.86$ & $1.91 \pm 0.35$ & $3.56 \pm 0.81$ \\
\hline
\end{tabular}
\label{table:mhi-mb}
\vskip 0.1in
The errors on the velocity-integrated \hii\ line luminosities in the penultimate column are measurement errors. The errors on the \hi\ mass in the last column are based on jackknife resampling.
\end{center}
\end{table*}

\section{The $\MHI- \mb$ scaling relation at $z \approx 0.35$} \label{sec:scaling}

\begin{figure*}[t]
\centering
\includegraphics[scale=0.9, trim={0.5cm 0.5cm 0 0}, clip]{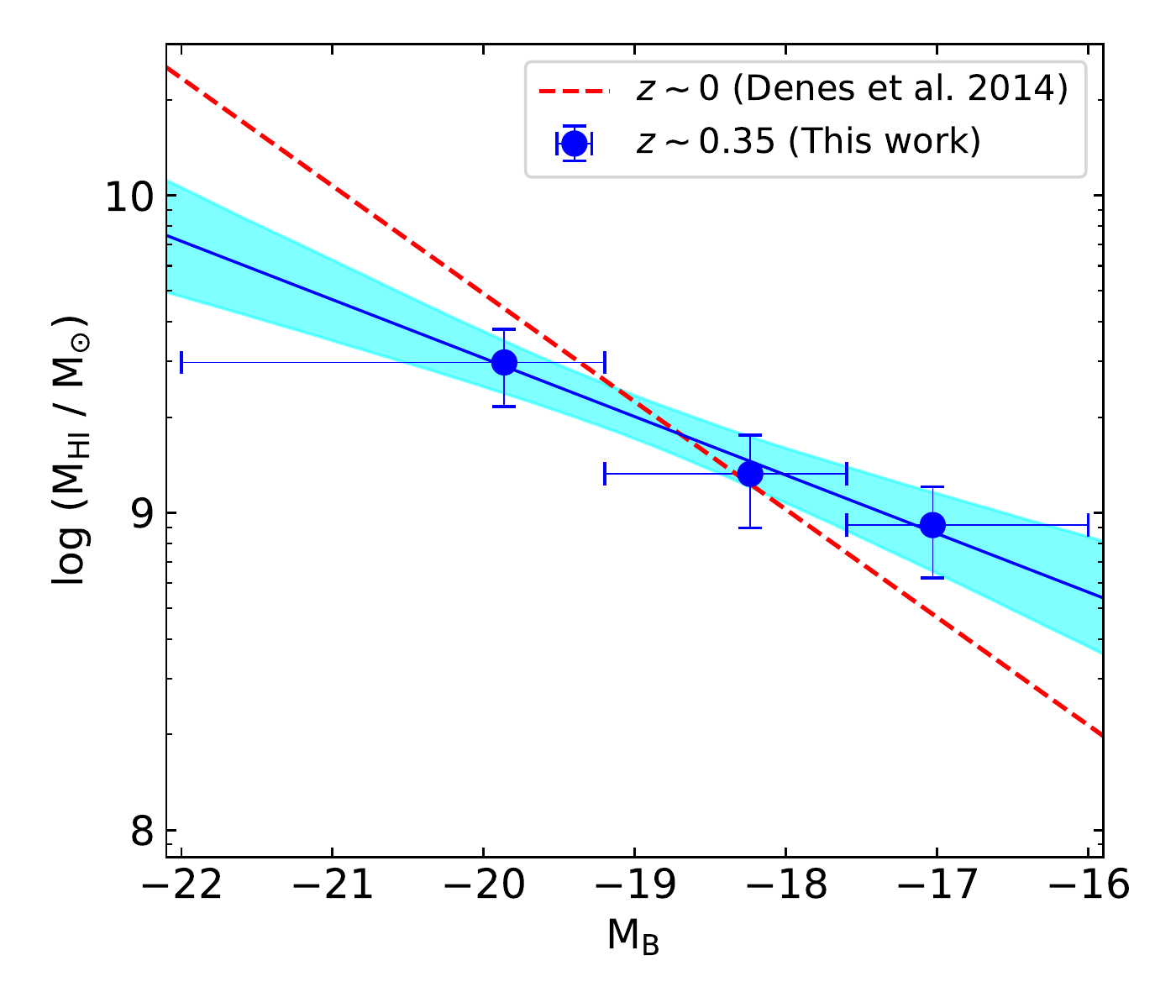}
\caption{\textbf{The $\mathbf {M_{HI}-M_B}$ relation for blue galaxies at $\mathbf{z\sim 0.35}$:} The solid blue circles show our estimates of $\rm \langle log (\MHI)\rangle$ values in blue galaxies in the three  $\mb $ bins at $z\approx 0.35$ plotted against the median $\mb$ values. The error bars are estimated by jackknife re-sampling. The horizontal error bars show the bin widths. The solid black line shows the best-fit $\MHI - \mb$ scaling relation at $z \approx 0.35$, while the blue shaded region shows the 68\% confidence interval around the scaling relation. The red dashed line indicates the $\MHI - \mb$ relation at $z \approx 0$ \citep{denes14mnras}.}
\label{fig_2}
\end{figure*}

For galaxies in the local Universe, the dependence of the \hi\ mass with the absolute B-band luminosity is well described by a linear relation between the logarithm of the \hi\ mass, $\rm log[\MHI]$, and the absolute B-band magnitude \citep[e.g.][]{denes14mnras}. The best-fit scaling relation can be directly obtained from the measurements of the \hi\ masses of individual galaxies.  The local $\MHI - \mb $ scaling relation (for late-type galaxies) is well described by the parametric form 
\begin{equation}
{\rm log} \left( \frac{\mh }{{\rm M}_{\odot}} \right) = \alpha + \beta \times \mb
\label{eqn:denes}
\end{equation}
where $\beta = -0.34 \pm 0.01$ and $\alpha = 2.89 \pm 0.11$ \citep{denes14mnras}. The scaling relation has a logarithmic scatter of $\sigma = 0.26$~dex \citep{denes14mnras}. 

The $\MHI - \mb$ scaling relation of Equation~\ref{eqn:denes} yields the average value of $\rm log[\MHI]$, i.e. $\langle \rm log[\MHI] \rangle$, at different values of $\mb$. However, in the present analysis, the method of \hii\ stacking yields the average value of $\MHI $, i.e. $\langle \MHI \rangle$, for each $\mb$ bin, and we can thus directly obtain $\rm log[ \langle \MHI \rangle]$, rather than $\langle \rm log[\MHI] \rangle$. The relation between $\rm log[\langle \MHI \rangle]$ and $\langle \rm log[\MHI] \rangle$ in an $\mb$ bin depends on the distribution of the \hi\ masses of the galaxies in that bin. Assuming that the \hi\ masses of the galaxies within a bin are log-normally distributed, we have
\begin{equation}
\langle \rm log [\MHI] \rangle \;\;= \;\;  log [\langle \MHI \rangle] - \frac{{\rm ln} \: 10}{2} \sigma^2  \; ,
\label{eqn:mhiave}
\end{equation}
where $\sigma$ is the logarithmic scatter of the \hi\ masses in the bin, with $\sigma = 0.26$~dex at $z \approx 0$ \citep{denes14mnras}. 
However, there exists no estimate of the scatter of the $\MHI - \mb$ scaling relation at $z \sim 0.35$. We hence assume that galaxies at $z \approx 0.35$ show the same scatter as galaxies in the nearby Universe, and use the value $\sigma = 0.26$ to estimate $\langle \rm log [\MHI] \rangle$ for different $\mb $ bins. Later in this section, we show that $\sigma = 0.26$ is consistent with the constraints on the scatter obtained from our full sample of galaxies. 

To measure the $\mh - \mb $ scaling relation for galaxies at $z\sim 0.35$, the values of $\langle {\rm log}\: \mh \rangle$ in the three $\mb $ bins of Table~\ref{table:mhi-mb} were estimated from Equation~\ref{eqn:mhiave} by subtracting 0.078~dex from the measurements of ${\rm log}[\langle \MHI \rangle]$. We note that this assumes a log-normal distribution of the individual \hi\ masses with a scatter of $\approx 0.26$~dex in each $\mb$ bin. A scaling relation of the parametric form\footnote{We have chosen to normalize the scaling relation at the centre of the $\mb $ range of our sample, $\mb = -19$, so as to obtain the lowest covariance between the errors on the slope and the normalization of the relation.} 
\begin{equation}
{\rm log} \left( \frac{\mh }{{\rm M}_{\odot}} \right) = \alpha_{19} + \beta (\mb + 19)
\label{eqn:mhi-mb-form}
\end{equation}
was then fitted to the estimates of $\langle {\rm log}[\MHI] \rangle$ at the median $\mb$ values of Table~\ref{table:mhi-mb}, via a chi-square minimization approach. The uncertainties associated with the best-fit parameters were estimated from the error covariance matrix of the fit, obtained using the jackknife standard errors as $1\sigma$ errors on the data points. The best-fit $\mh - \mb$ scaling relation for blue galaxies at $z \approx 0.35$ is found to be
\begin{equation}
{\rm log} \left(\frac{\MHI}{\msun}\right)  = (9.303 \pm 0.068) - (0.184 \pm 0.053)(\mb + 19) \,.
\label{eqn:mhi-mb-new}
\end{equation}

Figure~\ref{fig_2} shows the $\MHI - \mb$ scaling relation of Equation~\ref{eqn:mhi-mb-new}, with, for comparison, the local $\MHI - \mb$ relation of \citet{denes14mnras}. It is clear that the $\MHI - \mb$ relation at $z \approx 0.35$ is significantly flatter than the local relation, with the slopes being discrepant at $\approx 3\sigma$ significance. Our results indicate that luminous galaxies at $z \approx 0.35$ are \hi-poor on average, compared to their local counterparts, while fainter galaxies at $z\approx 0.35$ tend to contain more \hi\ than similar galaxies at the present epoch.

For a symmetric logarithmic scatter around the $\MHI - \mb$ scaling relation, $\langle {\rm log}[\mh] \rangle = {\rm log}[\MHI^{\rm med}]$, where $\MHI^{med}$ is the median value of $\MHI$. This implies that {\it median-stacking} of the \hii\ emission signals from the galaxies in each $\mb$ bin can be used to directly estimate $\langle {\rm log}[\mh] \rangle$. However, in the current analysis, the detection significances of the median \hi\ masses in the different $\mb$ bins are too low to use this approach. We hence used mean-stacking (which yields a higher, $> 4\sigma$, detection significance in each $\mb$ bin) and Equation~\ref{eqn:mhiave} to estimate $\langle {\rm log}[\MHI] \rangle$. If the median \hi\ masses in multiple $\mb$ bins could be measured with high significance, median-stacking would be a better approach to robustly determine the \hi\ scaling relations. 

While the sensitivity of the current \hi\ stacks is not sufficient to detect the median \hi\ mass at sufficient significance in each $\mb$ bin, we do detect the median \hii\ emission signal of the full sample of galaxies with a high significance (see Section~\ref{sec:stacking}). We obtain a median \hi\ mass of $(1.17 \pm 0.19) \times 10^9\ \msun $, while the mean \hi\ mass of the full sample is $(1.57 \pm 0.22) \times 10^9\ \msun $. We use these estimates of the mean and the median \hi\ masses of the full sample to constrain the scatter around the $\MHI - \mb$ scaling relation. For a log-normal distribution for the \hi\ masses of the full sample (i.e. effectively neglecting the slope of the scaling relation), the log-normal scatter can be estimated from Equation~\ref{eqn:mhiave}, as
\begin{equation}
\sigma_{\rm FS} = {\rm \sqrt{(2/ ln\ 10)\ log [\langle \MHI \rangle / \MHI^{med}]}}  \; ,
\label{eqn:scatter}
\end{equation}
where we have used the fact that, for a log-normal distribution, $\rm \langle log [\MHI] \rangle = log [\MHI^{med}]$. For our full sample, we find $\rm \sigma_{FS} = 0.33 \pm 0.10$~dex. However, this is an upper limit to the true scatter in the \hi\ mass around the scaling relation, because a non-zero slope in the relation causes an additional spread in the \hi\ mass, over and above the log-normal scatter.
We hence used a Monte Carlo approach to estimate the true scatter, combining the measured value of $\rm \sigma_{FS} = 0.33 \pm 0.10$~dex with the known $\mb $ distribution of our sample. We randomly assigned \hi\ masses to galaxies in our sample based on their $\mb $ values and the measured slope of the scaling relation\footnote{Note that the normalization of the scaling relation has no effect on this analysis.}, assuming different values of the true scatter, and measured $\rm \sigma_{FS}$ for each distribution using Equation \ref{eqn:scatter}. For the allowed range  $\rm \sigma_{FS} = 0.33 \pm 0.10$~dex, we obtain a true scatter of $\sigma = 0.21^{+0.14}_{-0.21}$~dex. We note that this estimate of the true scatter is entirely consistent with the local value, $\sigma = 0.26$~dex. 


Finally, the DEEP2 and DEEP3 galaxy samples are complete to $\rm R_{AB} < 24.1$, corresponding to $\mb < -18$ for our redshift range. This implies that the faintest $\mb$ bin, with $-18 < \mb \leq -16$, suffers from incompleteness \citep{berainprep}, with the $\mb$ distribution of galaxies in this bin different from that of the true cosmological distribution at $z \approx 0.35$. To test whether this incompleteness might introduce a bias in the inferred parameters of the $\MHI - \mb$ scaling relation, we carried out a Monte Carlo simulation to test whether a given input scaling relation would be recovered via our approach, given the actual $\mb$ distribution of our galaxy sample. This was done by assuming a range of $\MHI - \mb$ scaling relations of the form of Equation~\ref{eqn:denes}, each with a logarithmic scatter of 0.26~dex. For each such scaling relation, we assigned each galaxy in our sample a value of $\MHI$ based on its $\mb$ value. We then measured the average \hi\ mass in each of the three $\mb$ bins of Table~\ref{table:mhi-mb}, determined the value of $\langle {\rm log}[\MHI] \rangle$ for each bin, and carried out a least-squares fit to determine the best-fit $\MHI - \mb$ scaling relation. The incompleteness of the $\mb$ sample was found to have no effect on the normalization of the $\MHI - \mb$ relation; we hence fixed the value of $\alpha_{19}$ to $9.30$. The slope of the input scaling relation, $\beta$, was allowed to vary between $-0.34$ and 0, i.e. between the value in the local Universe and a flat $\MHI - \mb$ relation. For all values of $\beta$ in this range, the obtained best-fit scaling relation was found to be consistent with the input scaling relation, within the measurement uncertainties. We hence conclude that the $\mh - \mb $ scaling relation estimated in this study, for which the $\beta$ value lies within the simulated range, is not significantly affected by the incompleteness of the galaxy sample.

\section{The \hi\ mass function at $z \sim 0.35$} \label{sec:h1mf}

\subsection{Estimating the \himf: Validation in the local Universe} \label{subsec:h1mfestimate}

\begin{figure*}[t]
\begin{center}
\includegraphics[scale=0.65, trim={0.6cm 0.5cm 0.5cm 0cm}, clip]{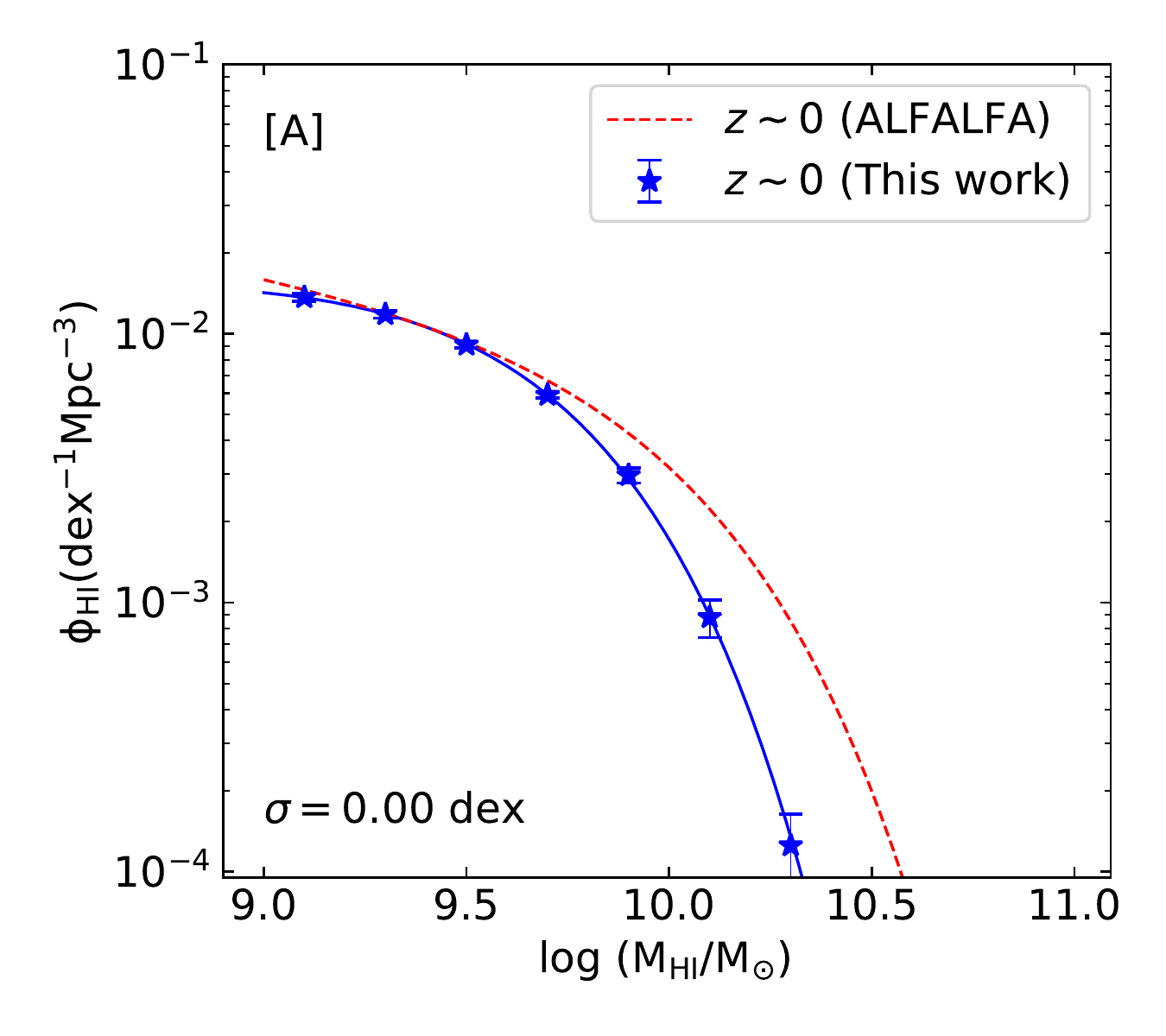}
\includegraphics[scale=0.65, trim={2.5cm 0.5cm 0.5cm 0cm}, clip]{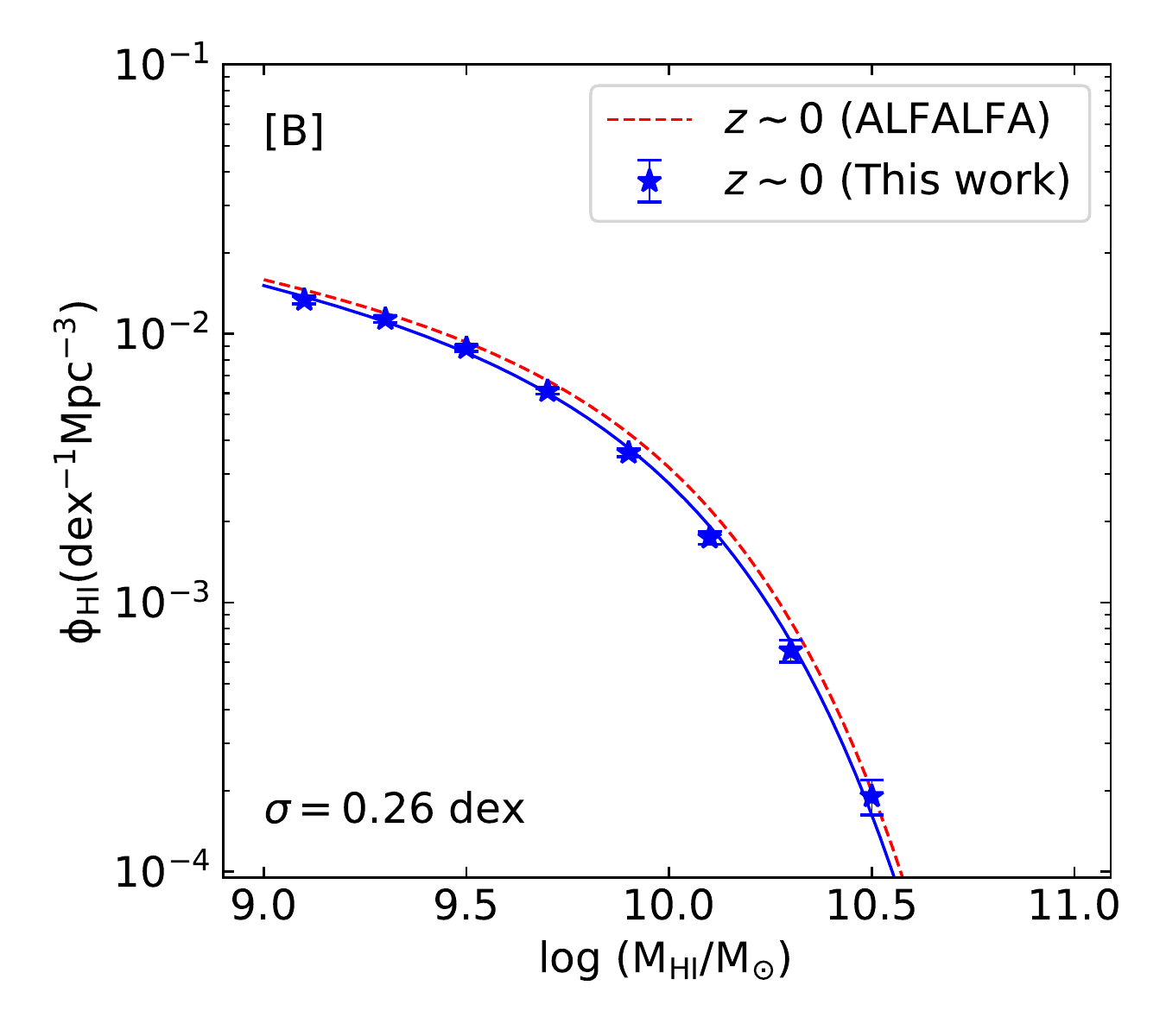}
\caption{The \himf\ at $z \approx 0$ inferred by combining the local $\MHI - \mb$ relation with the B-band luminosity function, with [A]~no assumed scatter in the $\MHI - \mb$ relation, and [B]~a scatter of $\sigma = 0.26$~dex, equal to that in the local Universe \citep{denes14mnras}. In each panel, the blue stars show the \himf\ of local galaxies inferred here, while the solid line shows the best-fit Schechter function to the inferred \himf. The dashed line indicates the  \himf\ measured from the unbiased ALFALFA survey \citep{jones18mnras}. It is clear that the assumption of zero scatter results in a significant under-estimate of the number of high-mass galaxies, while the assumption $\sigma=0.26$~dex yields a \himf\ that is in good agreement with the directly measured one.}
\label{fig:himf-local}
\end{center}
\end{figure*}

\begin{figure*}[t]
\begin{center}
\includegraphics[scale=0.65, trim={0 0.5cm 0 0}, clip]{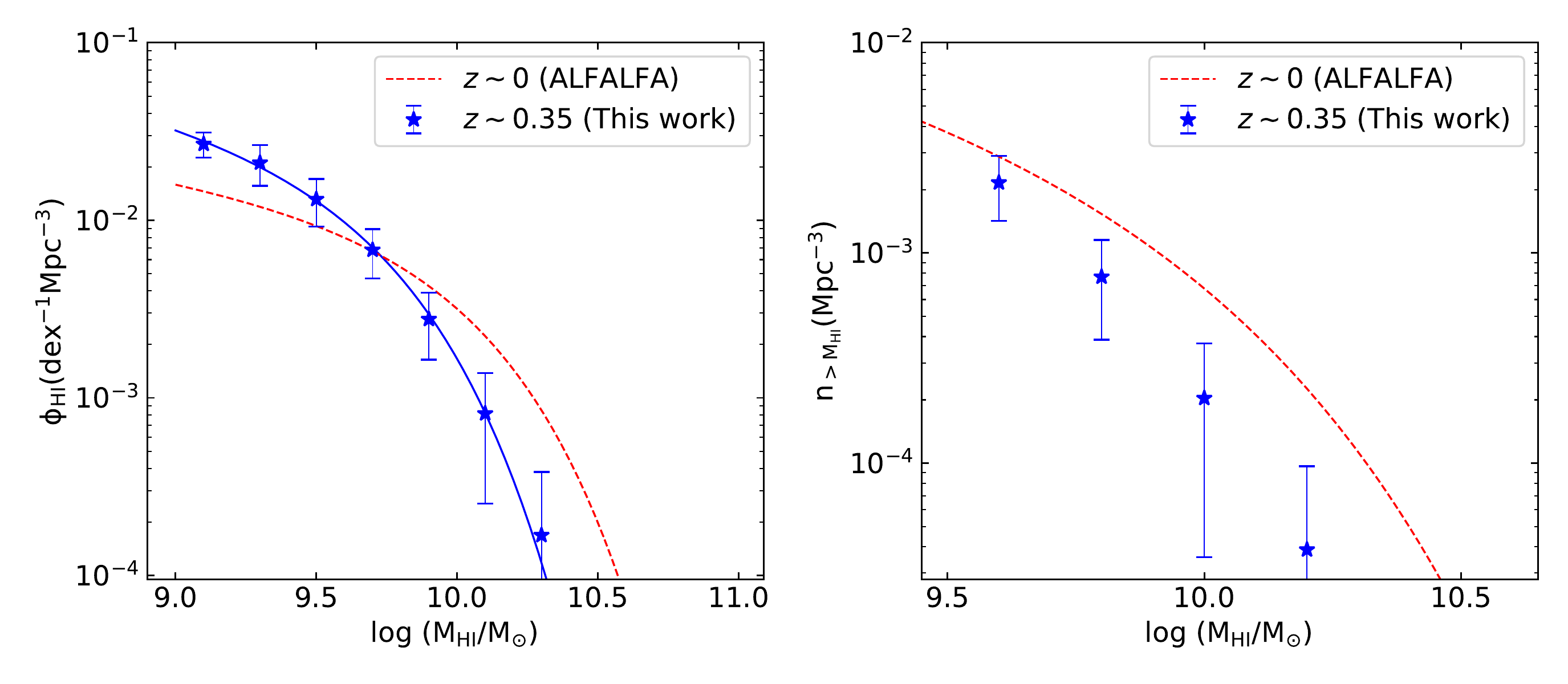}
\caption{[A]~The \himf\ at $z \approx 0.35$, inferred by combining our $\MHI - \mb$ relation with the ALHAMBRA B-band luminosity function \citep{lopezsanjuan17aa}. The blue stars represent the inferred \himf\ of blue galaxies at $z \sim 0.35$, while the dashed line shows the ALFALFA \himf\ at $z \approx 0$ \citep{jones18mnras}. [B]~The inferred number density of blue star-forming galaxies at $z \approx 0.35$ above a given \hi\ mass, as a function of $\rm log (\mh)$. The dashed curve shows the same quantity in the local Universe, from the \himf\ of ALFALFA survey \citep{jones18mnras}.}
\label{fig:himf-highz}
\end{center}
\end{figure*}

The \himf, $\Phi (\mh)$, is defined as the number of galaxies within a unit comoving cosmic volume having \hi\ masses between $\mh$ and $\mh +{\rm d} \mh$ for a given \hi\ mass, $\mh$. This can be estimated either via unbiased \hii\ emission surveys \citep[e.g.][]{zwaan05mnras,jones18mnras}, or by combining the $\mh - \mb $ scaling relation with the B-band galaxy luminosity function  \citep[e.g.][]{rao93apj, zwaan01mnras}. The early studies that followed the latter approach were analytical in nature, neglecting the scatter in the $\MHI - \mb$ relation. However, the scatter in this relation is critical to correctly estimate the \himf: Ignoring the scatter leads to  under-estimating the \himf\ at the high-mass end, as we demonstrate below (see Fig.~\ref{fig:himf-local}).

We followed a Monte Carlo approach to estimate the \himf,  combining the B-band  luminosity function and the $\MHI - \mb$ scaling relation. We initially applied this approach to galaxies in the local Universe, where the \himf\ has been directly measured from unbiased \hii\ surveys. The B-band luminosity function at $z \approx 0$ is well described by the Schechter parametrization \citep[e.g.][]{schechter76apj, faber07apj}
\begin{equation}
\phi (L_{\rm B}) = \left( \frac{\phi_*}{L_{\rm B}^*}\right) \left( \frac{L_{\rm B}}{L_{\rm B}^*} \right)^{\alpha} e^{-L_{\rm B}/L_{\rm B}^*} \;,
\label{eqn:b-band}
\end{equation}   
with $\alpha = -1.03 $, $\phi_* = 5.9 \times 10^{-3}\:{\rm Mpc^{-3}}$ and $M_{\rm B}^* = -20.04$ \citep{bell03apjs, faber07apj}.

Using the above B-band luminosity function, we generated a sample of galaxies with known absolute B-band magnitudes within a cosmic volume of $\rm 10^7\: {\rm Mpc}^3$. We assigned each galaxy a value of $\rm log (\mh)$ via the local $\mh - \mb $ relation of Equation~\ref{eqn:denes} \citep{denes14mnras}. This 
was done via two approaches, [A]~assuming no scatter in the $\MHI - \mb$ relation, and [B]~assuming a logarithmic scatter of 0.26~dex, as measured in the local Universe \citep{denes14mnras}. Finally, for each approach, we inferred a non-parametric \himf\ by counting the number of galaxies in logarithmic bins of \hi\ mass (with a bin width of 0.2~dex), and dividing this number by the width of the bin and by the total cosmic volume. The errors on the inferred \himf\ were estimated from $10^4$ realizations of the galaxy sample.

 
Figures~\ref{fig:himf-local}[A] and [B] show the inferred local \himf\ from our approach, for the cases with zero assumed scatter and $\sigma = 0.26$~dex, respectively, along with the direct estimate of the local \himf\ from the unbiased ALFALFA survey \citep{jones18mnras}. We note the $\mh - \mb $ scaling relation was measured for galaxies with $\mh > 10^9\ \msun$, and hence restrict the figures to galaxies above this \hi\ mass. It is clear from Fig.~\ref{fig:himf-local}[A] that neglecting the scatter in the $\MHI - \mb$ relation results in significantly under-estimating the \himf\ at the high-mass end. Conversely, Fig.~\ref{fig:himf-local}[B] shows that using a scatter of $0.26$~dex yields an \himf\ that is in excellent agreement with the ALFALFA \himf. 

Following the literature, we assume the following Schechter function form for the \himf\
\begin{equation}
\Phi (\mh) = \frac{\Phi_*}{M_0^*} \left( \frac{\mh}{M_0^*} \right)^a exp \left[-\left(\frac{\mh}{M_0^*}\right) \right] \;.
\label{eqn:himf-local}
\end{equation}
The best-fit Schechter parameters for our inferred \himf\ are $\Phi_* = (4.0 \pm 0.4) \times 10^{-3}\:{\rm Mpc^{-3}}$, $a  = -1.29 \pm 0.03$, and ${\rm log}(M_0^*/M_{\odot}) = 9.94 \pm 0.03$. These are all in good agreement with the ALFALFA estimates of the same parameters, $\Phi_* = (4.5 \pm 0.2) \times 10^{-3}\:{\rm Mpc^{-3}}$, $a  = -1.25 \pm 0.02$, and ${\rm log}(M_0^*/M_{\odot}) = 9.94 \pm 0.01$ \citep{jones18mnras}. 

We conclude that it is indeed possible to determine the \himf\ by combining the B-band luminosity function with the $\MHI - \mb$ relation, but that a critical ingredient in the estimate is the scatter in the latter relation. Not including this scatter results in an under-estimate of the \himf\ at the high-mass end, while over-estimating the scatter yields an over-estimate of the number of high-mass galaxies. Thus, an accurate estimate of the scatter in the $\MHI - \mb$ relation is crucial for an estimate of the \himf\ via this approach.

\subsection{Determining the \himf\ at $z \approx 0.35$} \label{subsec:h1mfz35}

Having confirmed that combining the B-band luminosity function and the $\MHI - \mb$ scaling relation yields the expected \himf\ at $z \approx 0$, we used the same approach to estimate the \himf\ at $z \approx 0.35$ for the blue galaxies of our EGS survey. For this, we used the B-band luminosity function of star-forming galaxies at $0.2 < z < 0.4$ from the ALHAMBRA survey\footnote{Similar results were obtained on using the B-band luminosity function of blue galaxies at $z \approx 0.35$ from the DEEP2 survey \citep{willmer06apj}.} \citep[$\alpha = -1.29 $, $\phi_* = 3.1 \times 10^{-3}\:{\rm Mpc^{-3}}$ and $M_{\rm B}^* = -20.85$;][]{lopezsanjuan17aa}. We note that the B-band luminosity function at $z\approx 0.3$ is different from that in the local Universe, with a higher number density of luminous galaxies, with $M_{\rm B}<-20$, at $z \approx 0.3$. The measured $\mh - \mb$ scaling relation from the present work (Equation~\ref{eqn:mhi-mb-new}) was used to assign \hi\ masses to the galaxies within a volume of $10^7 \: {\rm Mpc}^3$, again assuming a logarithmic scatter of 0.26~dex. The errors on the inferred \himf\ were estimated from $10^4$ realizations of the galaxy sample via a Monte Carlo approach, taking into account the errors on the parameters of the $\mh - \mb$ scaling relation of Equation~\ref{eqn:mhi-mb-new}.

Figure \ref{fig:himf-highz}[A] shows the inferred \himf\ for blue galaxies at $z \approx 0.35$; for comparison, the dashed line shows the ALFALFA \himf\ at $z \approx 0$. Note that we only considered galaxies with $\mb \leq -16$ in our analysis, the $\mb$ range over which we measured the $\MHI - \mb$ relation. The relatively large uncertainties on the inferred \himf\ do not allow us to obtain robust estimates of the best-fit Schechter parameters. However, Figure~\ref{fig:himf-highz}[A] clearly indicates that there are far fewer massive galaxies at $z \approx 0.35$, with $\MHI \gtrsim 10^{10} \; \msun$, than in the local Universe. Figure~\ref{fig:himf-highz}[B] shows the number density of galaxies above a given \hi\ mass at $z \approx 0.35$ (blue stars) and at $z \approx 0$ (dashed curve) as a function of $\rm log[\mh ]$. The number density of blue galaxies\footnote{Note that the uncertainties associated with the number density at $z \approx 0$ are dominated by the conservative estimate of cosmic variance of \citet{jones18mnras}, obtained by comparing results from the ALFALFA Spring and Fall samples.} with $\MHI > 10^{10} \; \msun$ is $(6.8 \pm 1.2) \times 10^{-4} \; {\rm Mpc}^{-3}$ at $z \approx 0$ \citep[assuming the ALFALFA \himf;][]{jones18mnras} but only $(2.0 \pm 1.7) \times 10^{-4} \; {\rm Mpc}^{-3}$ at $z \approx 0.35$. We thus find evidence that the cosmological number density of galaxies with high \hi\ masses, $\MHI > 10^{10} \; \msun$ at $z \approx 0.35$ is smaller by a factor of $\approx 3.4$ than in the local Universe. Finally, we note that Figure~\ref{fig:himf-highz}[A] also indicates that there are more ``low-mass'' galaxies with \hi\ masses $\approx 10^9 \; \msun$ at $z \approx 0.35$ than in the local Universe. Interestingly enough, these results are both broadly consistent with predictions from the GALFORM semi-analytical model \citep{baugh2019mnras}.


The fact that the cosmic number density of galaxies with a high \hi\ mass, $\mh \gtrsim 10^{10 } \; \msun$, increases significantly from $z \approx 0.35$ to the present epoch is directly linked to the relatively low \hi\ content of luminous galaxies at $z \approx 0.35$ compared to the local Universe (see Section~\ref{sec:scaling}). Our results indicate that massive,  luminous galaxies at $z \approx 0$ have acquired a significant amount of \hi\ over the past four Gyr, through either merger events or accretion from the CGM. 

It should be emphasized that the cosmic volume of the EGS covered in our uGMRT observations, $\approx 4.7 \times 10^4$~comoving~Mpc$^3$, is relatively small. Our results could hence be affected by cosmic variance \citep[e.g.][]{driver10mnras}. For example, some of our galaxies may reside in dense environments like groups or clusters \citep[e.g.][]{gerke12apj}, which could affect their \hi\ content \citep[e.g. via ram-pressure stripping;][]{gunn72apj,chung09aj}. We have found no direct evidence for the presence of galaxy clusters within our target volume, but cannot rule out the possibility of galaxy groups. 
Similar studies covering a larger cosmic volume would be useful to test the results of the present survey.

\section{Summary} \label{sec:conclusion}

We report the first measurement of the $\MHI - \mb$ scaling relation at cosmological distances, $z \approx 0.35$, using the technique of spectral line stacking applied to a deep GMRT \hii\ emission survey of the EGS at $z \approx 0.2-0.42$. We obtain the relation ${\rm log} \left(\MHI/\msun \right)  = (9.303 \pm 0.068) - (0.184 \pm 0.053)(\mb + 19)$, for blue star-forming galaxies at $z \approx 0.35$, significantly flatter than the corresponding relation in the local Universe. This implies that faint star-forming galaxies at $z \approx 0.35$ have a higher \hi\ mass than similar galaxies at $z \approx 0$, while luminous galaxies have a lower \hi\ mass than their local counterparts. We combined this $\MHI - \mb$ scaling relation (with a scatter assumed to be equal to that in the local Universe) with the known B-band luminosity function of star-forming galaxies at $z = 0.2-0.4$ to determine the \hi\ mass function at $z \approx 0.35$. We find that the cosmological number density of galaxies with $\MHI \gtrsim 10^{10} \; \msun $ at $z \approx 0.35$ is smaller by a factor of $\approx 3.4$ than at the present epoch. Conversely, there appear to be more galaxies with low \hi\ masses, $\MHI \approx 10^9 \ \msun$, at $z \approx 0.35$ than at $z \approx 0$. While the comoving volume covered by our survey is relatively small, implying that cosmic variance may be an issue, our results indicate that luminous, massive galaxies in the local Universe have acquired a significant amount of \hi\ over the last $\approx 4$~Gyr, via either mergers or accretion.

\begin{acknowledgments}
We thank the staff of the GMRT who have made these observations possible. The GMRT is run by the National Centre for Radio Astrophysics of the Tata Institute of Fundamental Research. AB and NK thank Aditya Chowdhury and Balpreet Kaur for detailed comments on an earlier draft of the manuscript and many discussions on \hii\ stacking that have contributed to this paper. NK acknowledges support from the Department of Science and Technology via a Swarnajayanti Fellowship (DST/SJF/PSA-01/2012-13). AB, NK, $\&$ JNC also acknowledge the Department of Atomic Energy for funding support, under project 12-R\&D-TFR-5.02-0700. 
\end{acknowledgments}
\bibliographystyle{aasjournal}

\end{document}